\documentclass[prd,showpacs,preprintnumbers,floatfix,superscriptaddress,10pt]{revtex4}
\usepackage[mathscr]{eucal}
\usepackage{color}
\usepackage[dvips]{graphicx}
\usepackage{epsf}
\usepackage{bm}

\newcommand{\be}{\begin{equation}}
\newcommand{\ee}{\end{equation}}
\newcommand{\ba}{\begin{eqnarray}}
\newcommand{\ea}{\end{eqnarray}}

\begin{document}

\title{Bulk viscosities of a cold relativistic superfluid: color-flavor locked quark matter}

\author{Massimo Mannarelli and Cristina Manuel}

\address{$^1$ Instituto de Ciencias del Espacio (IEEC/CSIC) \\
Campus Universitat Aut\` onoma de Barcelona,
Facultat de Ci\` encies, Torre C5 \\
E-08193 Bellaterra (Barcelona), Spain}

\begin{abstract}
We consider  the phonon contribution to the bulk viscosities $\zeta_1, \zeta_2$ and $\zeta_3$ of a cold relativistic superfluid.
We  assume the  low temperature $T$ regime and that the transport  properties of the system are dominated by the phonons. We use kinetic theory in the relaxation time approximation  and the low energy effective field theory
of the corresponding system.  The parametric dependence of the  bulk viscosity coefficients is fixed once
the equation of state is specified, and the phonon dispersion law to cubic order in momentum is known. We first present a general  discussion, valid for any superfluid, then we focus on the color-flavor locked superfluid because all the parameters
needed in the analysis can be  computed in the high density limit of  QCD, and also because of the possible astrophysical applications. For the three independent bulk viscosity coefficients we find that
they scale with the temperature as  $\zeta_i \sim 1/T$, and that
in the conformal limit only the third coefficient $\zeta_3$ is non-zero.
\end{abstract}

\maketitle

\section{Introduction}

Superfluidity was first discovered in $^4$He 
when this was cooled at temperatures below 2.17 K \cite{landaufluids,IntroSupe}. Later on  it was found that various quantum liquids, both bosonic and fermionic,  exhibit the same property.  The phenomenon is due to  the appearance of a quantum condensate, which spontaneously breaks a global $U(1)$ symmetry.

Relativistic superfluids
might be realized  in the interior of neutron stars where the temperature is low and the energy scale of the particles is sufficiently high.   In particular, in the inner crust of neutron stars the attractive interaction between neutrons can lead to the formation of a BCS condensate, the system then becomes superfluid.
Moreover, if deconfined quark matter is present in the core of neutron stars
it will very likely be in a color superconducting phase~\cite{reviews}.
Quantum Chromodynamics (QCD) predicts that
at asymptotically high densities  quark matter is in the color-flavor locked  phase 
(CFL)~\cite{Alford:1998mk}.
In this phase up, down and strange quarks of all three colors pair forming a difermion condensate that is antisymmetric in color and flavor indices.
 The  order parameter breaks the baryonic number $U(1)_B$
symmetry spontaneously, and therefore CFL quark matter is  a superfluid as well.

If superfluidity occurs in the interior of compact stars, it should be possible to find 
signatures of its presence through a variety of astrophysical phenomena. For example, the most natural 
explanation for the sudden spin-up of pulsars~\cite{Anderson:1975zze}, the so-called glitches, relies on the existence of
a superfluid component in the interior of the star, rotating much faster than the outer solid crust.
After the unpinning of the superfluid vortices, there is a transfer of angular momentum from the interior of the star to the outer crust, giving rise to the the pulsar glitch.

Another possibility to detect or discard the presence of relativistic superfluid phases 
consists in studying the evolution of the r-mode oscillations of  compact stars~\cite{Andersson:2000mf}. R-modes are non-radial oscillations of the star with the Coriolis force acting as the restoring force. They provide a severe limitation on the rotation frequency of the star through coupling to  gravitational radiation (GR) emission. When dissipative phenomena damp these r-modes the star can rotate without losing  angular momentum to GR. If dissipative phenomena are not strong enough,  these  oscillations  
will grow exponentially and the star will keep slowing down until some dissipation mechanism   is able to  damp the r-mode oscillations. Therefore,  the study of r-modes  is useful in  constraining  the stellar structure and can be used to rule out some specific matter phases.
For such studies it is necessary to consider in detail all the dissipative processes and to compute the corresponding  transport coefficients. 

On the theoretical ground, there is renewed interest in the topic of this manuscript, as it has been recently proposed an holographic
model to describe relativistic superfluidity \cite{Hartnoll:2008vx,Herzog:2008he}.
With the holographic techniques one intends to model a strongly interacting gauge theory with a weakly interacting gravity dual,
having in this way a prescription to study many properties of the gauge system, such as its transport
coefficients.

There are different formulations of the hydrodynamical equations governing  a relativistic superfluid when there is no 
dissipation~\cite{khalatnikov,Carter-Kha,Carter-Lang,Carter-Kha-annals,Son:2000ht,And-Com}. They were derived as
relativistic generalizations of  Landau's two-fluid model of non-relativistic superfluid dynamics~\cite{landaufluids,IntroSupe}. In the non-dissipative limit it is possible to show that all these  approaches are equivalent.
The dissipative terms which enter into the relativistic hydrodynamical equations of one the above approaches have been derived in Ref.~\cite{Gusakov:2007px}. As it occurs in the non-relativistic case, for a relativistic superfluid  one can define a thermal conductivity, $\kappa$,
a shear viscosity, $\eta$, and four bulk viscosity coefficients, $\zeta_1,\zeta_2,\zeta_3,\zeta_4$. While $\zeta_2$ has the same meaning
as the bulk viscosity of a normal fluid, $\zeta_1$, $\zeta_3$ and $\zeta_4$ refer to dissipative process  that lead to entropy production only in the presence of a space-time
dependent relative motion between the superfluid and the normal fluid components~\cite{landaufluids,IntroSupe}.

This paper is  devoted to determine the  phonon contribution to the  bulk viscosity
coefficients of a  cold relativistic superfluid. The superfluid phonon is the Goldstone mode associated with the spontaneous breaking
of a $U(1)$ symmetry. Being a massless mode, at low temperatures it gives the leading contribution to  both the thermodynamics and transport phenomena.  The contribution of other quasiparticle degrees of freedom, which typically have an associated non-zero
energy gap $E$, is  suppressed when $T \ll E$ and
we will assume that we are in such a temperature regime.
This work is somehow a follow up of Ref.~\cite{Escobedo:2009bh},
where similar computations were done for a non-relativistic superfluid,  focusing on ultracold Fermi
gases close to the unitary limit. 

We perform our computation in the
framework of  kinetic theory, with the use of the relaxation time approximation (RTA). 
 We present a general discussion regarding the phonon contribution to the transport coefficients for a generic relativistic superfluid. The basic ingredient in our derivation is  the low energy effective field theory associated to
the superfluid phonon,
 which is essentially dictated by symmetry considerations \cite{Son:2002zn,Son:2005rv}.
The effective Lagrangian  is presented as the typical  expansion in derivatives of the Goldstone field, and with
it one can derive the phonon dispersion law and all the scattering rates  that are needed for the
computation of the transport coefficients to a given accuracy. It turns out that 
the scaling laws  for  the transport coefficients  depend on the coefficients of the effective field theory, which have  to be matched with the microscopic theory. 

After the presentation of the  discussion of  the bulk viscosities for a generic relativistic superfluid, in the last part of the present paper
we focus on the color-flavor locked  superfluid. 
At asymptotically high density, and thus weak coupling limit, all the coefficients of the  phonon effective field theory can be 
computed from QCD. 
A  study of the shear viscosity and of the bulk viscosity coefficient $\zeta_2$ due to phonons has been already presented in Refs.~\cite{Manuel:2004iv,Manuel:2007pz}. The contribution of kaons to $\zeta_2$  has  been studied in
Ref.~\cite{Alford:2007rw,Alford:2008}.  The contribution to thermal conductivity due to phonons was first estimated in Ref.~\cite{Shovkovy:2002kv}, 
and recently computed in Ref.~\cite{Braby:2009dw}.

This paper is organized as follows. In Section~\ref{hydro} we review the hydrodynamical equations for a
relativistic superfluid, when effects of dissipation are included. In Section~\ref{EFF} we present the
effective field theory for the  superfluid phonon, at the leading and 
next-to-leading order, following the work of Ref.~\cite{Son:2002zn,Son:2005rv}.
 In Section~\ref{pho-bulk} we derive the expressions for the phonon contribution
to the bulk viscosity coefficients. Based on these formulas, and on the scattering amplitudes for the pertinent processes,  we present a general discussion on the scaling
behavior of these transport coefficients in Section~\ref{Scalingbulk}. Section~\ref{bulkcfl} is devoted to  the study of  the CFL superfluid. We present our conclusions in  Section~\ref{conclu}.
We leave for  the Appendix  the proof that one needs the phonon dispersion law beyond linear order
to obtain the first non-trivial  contribution to the bulk viscosities.
Throughout we use natural units, $\hbar = c = k_B=1$, and metric conventions
$(+,-,-,-)$.

\section{Hydrodynamics of a relativistic superfluid}
\label{hydro}
The hydrodynamical equations of a relativistic superfluid have been derived using
different formulations \cite{khalatnikov,Carter-Kha,Carter-Lang,Carter-Kha-annals,Son:2000ht,And-Com}.
 Here we will use the one derived by Son~\cite{Son:2000ht}.
The  superfluid properties of a system 
arise from the  spontaneous breaking of a continuous $U(1)$ symmetry, with
 the appearance of a Goldstone mode. Since hydrodynamics
is an effective field theory valid at long time and long length scales,
 the standard fluid variables couple to the Goldstone field.

The hydrodynamical equations for the superfluid take the form of conservation laws
for both the current, $n^\mu$, and energy-momentum tensor, $T^{\mu \nu}$,  of the system
\begin{equation}
\partial_\mu n^\mu = 0 \ , \qquad \partial_\mu T^{\mu \nu} = 0 \ .
\end{equation}
One further adds the Josephson equation, which describes the dynamical evolution of
the Goldstone field, $\varphi$, or phase of the condensate
\begin{equation}
u^\mu \partial_\mu \varphi+ \mu = 0  \  ,
\end{equation}
where
 $\mu$ is the chemical potential of the system~\cite{footnote}.

In this formulation of the superfluid hydrodynamics there is a clear interpretation
of both the current and the stress-energy tensor  as being due to the sum of the normal fluid
part and  the coherent motion of the condensate (the superfluid), as  these are expressed 
 as
 \begin{eqnarray}
 T^{\mu \nu} & = & \left(\rho + P\right)u^\mu u^\nu -
P g^{\mu \nu}
+ V^2 \partial^\mu \varphi \partial^\nu \varphi  \  ,\\
 n^\mu & = & n_0 u^\sigma - V^2 \partial^\mu \varphi \  ,
\end{eqnarray}
where $u^\mu$ is the hydrodynamical velocity,  $\rho$ stands for the energy density, $P$ is the pressure, and $V$ is a variable proportional
to the quantum condensate. The energy density obeys the relation $\rho= ST + n_0 \mu - P$,
where $S$ is the entropy of the system.

The dissipative terms associated to the above hydrodynamical equations have been constructed in~\cite{Gusakov:2007px},
showing that in the non-relativistic limit they correspond to those appearing in Landau's two-fluid model. To this
end it is better to write  the hydrodynamical equations in terms of the new
 variable
\begin{equation}
w^\mu = - \left(\partial^\mu \varphi + \mu u^\mu \right) \ .
\end{equation}
Then, one can show that in the non-relativistic
limit the spatial component of this four vector becomes the counterflow velocity 
of Landau's equations, ${\bf w} =  m ({\bf v_s} -{\bf v_n})$,
with the identification ${\bf v_s} = -\frac{\nabla \varphi}{m}$ \cite{Gusakov:2007px}.

In order to construct the dissipative terms in the fluid equations one has to define a comoving frame. One
possible choice is that in the frame where $u_\mu = (1,0,0,0)$ 
the particle current and energy are fixed as
\begin{equation}
n^\mu = (\bar n, V^2 {\bf \nabla} \varphi)  \ , \qquad T^{00} = \epsilon \ .
\end{equation}
Now one imposes some restrictions to the possible dissipative terms that
might be added to the current and energy-momentum tensor, $n_d^\rho$ and $T^{\mu \nu}_d$, respectively. In 
particular, one can choose constraints similar to those of the Eckart frame for the normal fluid hydrodynamics,
$ n^\rho_d = 0$, and $ u_\mu u_\nu T^{\mu \nu}_d =0 \ $.

Dissipative terms are taken into account as follows. 
First, one modifies the Josephson equation
\begin{equation}
u^\mu \partial_\mu \varphi= - \mu- \chi \ ,
\label{josephson}\end{equation} 
with
\begin{equation}
\chi  =  -\zeta_3 \partial_\mu (V^2 w^\mu) - \zeta_4 \partial_\mu u^\mu \ .
\end{equation}
One also adds  dissipative terms to the energy-momentum tensor, 
which take the form 
\begin{eqnarray}
T^{\mu \nu}_d & = &   \kappa \left( \Delta^{\mu \gamma} u^\nu + \Delta^{\nu \gamma} u^\mu
\right) \left(\partial_\gamma T + T u^\delta \partial_\delta u_\gamma \right)  
+ \eta \Delta^{\mu \gamma}\Delta^{\nu \delta} \left(\partial_\delta u_\gamma + \partial_\gamma u_\delta + \frac 23 g_{\gamma \delta}
\partial_\alpha u^\alpha\right) \\
\nonumber
&+& \Delta^{\mu \nu} \left(\zeta_1 \partial_\gamma (V^2 w^\gamma) +\zeta_2 \partial_\gamma u^\gamma \right) \ ,
\end{eqnarray}
where
$\Delta^{\alpha \beta} = g^{\alpha \beta} - u^\alpha u^\beta$.
In principle, more terms respectful with the
symmetries of the problem are possible, 
 but those are neglected, assuming that they are small. One simply retains six transport
coefficients, namely the shear viscosity
$\eta$, the  thermal conductivity $\kappa$, and  $\zeta_1,\zeta_2,\zeta_3,\zeta_4$ which are the four bulk viscosity coefficients. 
In the non-relativistic
limit the expressions of the bulk viscosity coefficients agree with those introduced by Khalatnikov~\cite{IntroSupe}, up to some mass factors~\cite{Gusakov:2007px}.

According to the Onsager symmetry principle \cite{Onsager,landaufluids,IntroSupe}  the transport coefficients satisfy the relation $\zeta_ 1 = \zeta_4$, while the requirement of positive entropy production imposes that 
$\kappa, \eta, \zeta_2, \zeta_3$   are positive and that  $\zeta_ 1^2 \leq \zeta_2 \zeta_3$.  

The friction forces due to bulk viscosities can be understood as drops, with respect to their equilibrium values, in the main driving forces acting on the normal and 
superfluid components. These forces are given by the gradients of the pressure $P$ and the chemical potential $\mu$. One can write in the comoving frame
\ba
P &=& P_{\rm eq} - \zeta_1 {\rm div}(V^2 {\bf w}) -\zeta_2  {\rm div} {\bf u}\,, \\
\mu &=& \mu_{\rm eq} - \zeta_3 {\rm div}(V^2 {\bf w}) -\zeta_1  {\rm div}{\bf u}\, ,
\ea 
where $P_{\rm eq}$ and $\mu_{\rm eq}$ are the equilibrium pressure and chemical potential. Note that the drops in the driving forces   proportional to $\zeta_1$ and $\zeta_3$ lead to entropy production only when ${\rm div}\,{\bf w \neq 0}$. Indeed in case ${\rm div}\,{\bf w = 0}$, the entropy production rate is given by~\cite{IntroSupe,Gusakov:2007px}
\be
R = \zeta_2 (  {\rm div} {\bf u})^2 \,.
\ee 
Therefore, when  ${\rm div}\,{\bf w = 0}$, in the Josephson equation (\ref{josephson}) one has that   $\chi = -\zeta_1 \partial_\mu u^\mu$, but such a term does not lead to dissipation.

\section{Effective field theory for the phonon of a relativistic superfluid}
\label{EFF}

All the phonon properties that are needed for the evaluation of the bulk viscosities
can be extracted from the low energy effective field theory associated to the
relativistic superfluid.
This has been constructed in Refs.~\cite{Son:2002zn,Son:2005rv}, and we review it
here.

The effective field theory for the Goldstone field is constructed as a power
expansion over derivatives and over  fields, $\sim \partial^n \varphi^m$,
 allowing only the terms that respect  the underlying symmetries
of the system one is considering.  The coefficients appearing in the low energy Lagrangian can be in principle
computed from the microscopic theory, through a standard matching procedure.

Let us review the form of the effective field theory.
Consider a relativistic system that experiences the spontaneous breaking
of a $U(1)$ symmetry. If $\Theta$ is the phase of the field that gets
an expectation value in its ground state
$\Psi = | \Psi | e^{-i \Theta}$, 
then  $\Theta$ is related to the Goldstone field $\varphi$ by the relation
$\Theta = \mu_0 t - \varphi$, 
where $\mu_0$ is the zero temperature chemical potential~\cite{footnote}, and $t$ is the time.

At the lowest order in $\Theta$, the Lagrangian must be a function of 
\begin{equation}
{\cal X} = \frac 12 g^{\mu \nu} \partial_{\mu} \Theta \partial_\nu \Theta \ ,
\end{equation}
where $g^{\mu \nu}$ is the metric tensor.
Since we are not interested in curved space-time we can rewrite \begin{equation}
{\cal X} = \frac 12 \left(\mu_0^2 - 2 \mu_0 \partial_0 \varphi + \partial_\nu \varphi 
\partial^\nu \varphi \right) \ .
\end{equation}

The Lagrangian for the Goldstone field at lowest order (LO) is expressed as
\begin{equation}
{\cal L}_{\rm LO} = P[\sqrt{2 {\cal X}}] \ ,
\end{equation}
where $P$ is the zero-temperature pressure of the system one is considering. The reason why the Lagrangian
takes this form is 
that the effective action of the theory at its minimum for constant classical field configurations
has to be equal to the pressure.

 One can expand the
pressure function around $\mu_0$ finding in this way the Lagrangian 
\begin{equation}
 \label{L-BGB}
 {\cal L}_{\rm LO}  = \frac 12 (\partial_0 {\tilde \varphi})^2 -
\frac{c_s^2}{2} (\partial_i {\tilde \varphi})^2 - g_3
\partial_0 {\tilde \varphi} (\partial_\nu {\tilde \varphi}  \partial^\nu {\tilde \varphi}) +
g_4(\partial_\nu {\tilde \varphi}  \partial^\nu  
{\tilde \varphi})^2 + \cdots\ ,
\end{equation}
where we have rescaled the Goldstone field so as to have a kinetic term
properly normalized. We have also neglected a piece proportional to $\partial_0 \tilde \varphi$, only relevant for
vortex configurations. The constants $c_s, g_3, g_4$ can be written in terms of different ratios
of derivatives of the pressure. In particular, $c_s$ is seen to
agree with the zero temperature speed of sound of the system
\begin{equation}\label{cs}
c_s^2 = \frac{1}{\mu_0} \left(\frac{\partial P}{\partial \mu_0}\right)  \left(\frac{\partial^2 P}{\partial \mu_0^2}\right)^{-1} \ ,
\end{equation} 
while the coupling constants are expressed as
\begin{equation}
\label{LO-coupl}
g_3 = \frac{1}{2 \mu_0} \left( \frac{\partial^2 P}{\partial \mu_0^2} - \frac 1\mu_0 \frac{\partial P}{\partial \mu_0} \right)  \left(\frac{\partial^2 P}{\partial \mu_0^2}\right)^{-3/2} \ , \qquad g_4 = \frac{1}{8 \mu_0^2} \left( \frac{\partial^2 P}{\partial \mu_0^2} - \frac 1\mu_0 \frac{\partial P}{\partial \mu_0} \right)  \left(\frac{\partial^2 P}{\partial \mu_0^2}\right)^{-2} \ .
\end{equation}

At next-to-leading order (NLO) there are four different structures, but two
of them vanish in flat space-time. For simplicity,  we
will not write them down. At NLO one has

\begin{equation}
\label{Lag-at-NLO}
{\cal L}_{\rm NLO} = \frac{f_1({\cal X})}{\mu_0^2} \partial_\nu {\cal X} \partial^\nu {\cal X} + f_2({\cal X}) (\partial_\nu \partial^\nu \Theta)^2 \ ,
\end{equation}
and we have re-defined the functions that appear in Ref.~\cite{Son:2005rv} so as to make $f_{1}$ and $f_2$ dimensionless
functions.

We will  mainly be concerned with the effect that the NLO terms have in modifying
the phonon dispersion law. To this end, it is enough to expand the
Lagrangian up to terms which are quadratic in $\varphi$. 
We assume that the functions $f_{1,2}$  that appear in Eq.~(\ref{Lag-at-NLO}) are analytical, 
and that they can be expanded in a Taylor series around $\mu_0^2$. Thus,
after the necessary re-scaling to have a normalized kinetic term, one finds
\begin{equation}
{\cal L}_{\rm LO} + {\cal L}_{\rm NLO} = \frac 12 (\partial_0 {\tilde \varphi})^2 -
\frac{c_s^2}{2} (\partial_i {\tilde \varphi})^2 + m_1
\left(\partial_\nu \partial^0 {\tilde \varphi} \partial^\nu \partial_0 {\tilde \varphi} \right) 
+ m_2 (\partial_\nu \partial^\nu
{\tilde \varphi})^2 + \cdots
\end{equation}
where
\begin{equation}
\label{NLO-cts}
m_1 =  \frac{f_1(\mu_0^2)}{\frac{\partial^2 P}{\partial \mu_0^2}} \ , \qquad m_2 =  \frac{f_2(\mu_0^2)}{\frac{\partial^2 P}{\partial \mu_0^2}} \ ,
\end{equation}
which leads to the following modification of the dispersion equation for the phonon
\begin{equation}
\omega^2 - c_s^2  q^2 + 2 m_1  \omega^2 (\omega^2 -q^2) +
2 m_2 (\omega^2 - q^2)^2 = 0 \ .
\end{equation}

Assuming that $\omega, q \ll \sqrt{\frac{\partial^2 P}{\partial \mu_0^2} }$, then one finds 
the dispersion law 
\begin{equation}
\label{cubic-law}
\epsilon_q = c_s q + B q^3 + \cdots
\end{equation}
where 
\begin{equation}
B = \left( c_s (1-c_s^2) m_1 - \frac{1}{c_s}(1-c_s^2)^2 m_2 \right) \ . \label{B}
\end{equation}

The NLO Lagrangian also introduces corrections to the phonon self-couplings, that we will not explicitly write down here
as they will only be necessary if we were interested in computing corrections to the leading order behavior
of the transport coefficients.

It is also interesting to discuss the conformal invariance of the phonon effective field theory. If the
macroscopic system is conformal invariant, this symmetry should be inherited by the low energy effective
field theory.  At LO, this information is encoded in the explicit form of the pressure
of the system, and correspondingly, in 
the speed of sound that should take the value of  $1/\sqrt{3}$. At NLO, conformal invariance
puts some constraints on the form of the functions $f_1$ and $f_2$, which can only
depend on powers of the ratio ${\cal X}/\mu_0^2$. If conformal invariance is not a symmetry of
the system, the functions $P, f_1$ and $f_2$ will depend on the scale responsible for the
breaking of that symmetry.

It is important to stress the  general character of the present discussion, 
valid for any relativistic superfluid, independent of the fact that the system is  strongly coupled. Indeed in the above derivation one only employs the symmetries of the system. As an application one may think to the neutron superfluid which is believed to exist in the interior of neutron stars, or to the CFL 
superfluid that we will treat more carefully in the last part of the present manuscript.

\section{Phonon contribution to the frequency dependent bulk viscosities}
\label{pho-bulk}

At very low temperatures phonons give the leading thermal contribution to all the thermodynamic
properties of the superfluid. In the hydrodynamic regime, phonons also give the leading contribution
to the transport coefficients entering into the two-fluid equations. Khalatnikov
developed the kinetic theory associated to these degrees of freedom  for 
a non-relativistic superfluid~\cite{IntroSupe}. One can also construct the transport theory
associated to these Goldstone mode excitations of a relativistic superfluid \cite{Mannarelli:2008jq,Popov:2006nc}, for example using
a gravity analogue model, showing that in the non-relativistic limit one recovers the theory of Khalatnikov \cite{Mannarelli:2008jq}.

For the computation of the bulk viscosity coefficients we will use the method
suggested by Khalantnikov for superfluid $^4$He, and  recently used for 
the cold Fermi gas in the unitarity limit~\cite{Escobedo:2009bh}. In this approach one studies 
the dynamical evolution of the phonon number density ${\cal N}_{\rm ph}$, and relates the variation of this quantity with the transport coefficients of the system.  It has been shown in Ref.~\cite{Escobedo:2009bh}, that this approach
is equivalent  to solving the phonon transport equation in
the relaxation time approximation. Note that with the RTA  one obtains the correct parametric behavior of the transport
coefficients, but one cannot fix with precision the numerical factor in front of these quantities. In order to obtain such a factor, it
is necessary to solve numerically the corresponding Boltzmann equation, a rather complicated task.
For the astrophysical applications we have in mind, getting the correct parametric behavior of all the
transport coefficients is good enough, as this fixes the time scales of different physical processes.

Consider a perturbation to the system that changes the  
number of phonons per unit volume ${\cal N}_{\rm ph}$.  Collisional
processes will tend to restore the equilibrium phonon number density.
The dynamical evolution of ${\cal N}_{\rm ph}$ can be extracted from the  transport equation obeyed by the phonon distribution function~\cite{Mannarelli:2008jq}.
If we further use the relaxation time approximation to deal with the collision
term, and assume that the speed of sound is constant, then we obtain that
\be
\partial_{\nu} (u^{\nu}  {\cal N}_{\rm ph}) =   - \frac{\delta {\cal N}_{\rm ph}}{ \tau_{\rm rel}} \ ,
\ee
where $\delta {\cal N}_{\rm ph}$  measures the departure of the phonon density 
from its equilibrium value.
The relaxation time is given by
\be
\frac{1}{\tau_{\rm rel}} \sim \frac{\Gamma_{\rm ph}}{{\cal N}_{\rm ph}} \, ,
\ee 
where $\Gamma_{\rm ph}$ is the rate of change of the number of phonons, to be specified later on.

From now on we will work in the comoving frame $u^\mu =(1,0,0,0)$ to simplify the treatment.
In equilibrium, the number density of phonons is a function of the total current density, $\bar n= n_0 + V^2 \mu_0$,  and 
of the entropy $S$. Since we are interested in  astrophysical applications we shall  assume that the perturbation is  periodic, so that there is a dependence on time of the sort
$\sim \exp{(i \omega_c t)}$, 
with typical frequencies  $\omega_c$, as measured in
the  comoving frame. For astrophysical applications   $\omega_c$ is of the order of the frequency of rotation of the star.

Since the phonon number density depends on the current density and on the entropy, we have that
\be
\delta {\cal N}_{\rm ph}  = \frac{\partial {\cal N}_{\rm ph}}{\partial \bar n}  \delta \bar n    + \frac{\partial {\cal N}_{\rm ph} }{\partial S} \delta S \, .
\ee
 The hydrodynamic equations take the form 
\ba
u^\mu \partial_\mu \bar n & = & - \bar n \partial_\mu u^\mu - \partial_\mu (V^2 \omega^\mu)  \ , \\
u^\mu \partial_\mu S & = & - S \partial_\mu u^\mu \ ,
\ea
and therefore  in the comoving frame one can rewrite the phonon number variation as
\be
\delta {\cal N}_{\rm ph}  = \frac{\tau_{\rm rel}}{1 - i \omega_c \tau_{\rm rel}} \left\{
 \left(  {\cal N}_{\rm ph} - \bar n \frac{\partial {\cal N}_{\rm ph}}{\partial \bar n} - S \frac{\partial {\cal N}_{\rm ph} }{\partial S}   
 \right) {\rm div}{\bf u} - \frac{\partial {\cal N}_{\rm ph}}{\partial \bar n} {\rm div}(V^2 {\bf w}) 
   \right \} \ .
\ee

At this point one studies how the variation of the phonon density alters the equilibrium values
of the pressure and chemical potential~\cite{IntroSupe}. In this way one can identify the different  bulk viscosity coefficients and
if we define 
\begin{equation}
\label{defIs}
I_1 = \frac{\partial {\cal N}_{\rm ph}}{\partial \bar n} \ , \qquad I_2 = {\cal N}_{\rm ph} -   S \frac{\partial {\cal N}_{\rm ph}}{\partial S} - \bar n \frac{\partial {\cal N}_{\rm ph}}{\partial \bar n} \ ,
\end{equation}
we obtain that 
\ba\label{xi1}
\zeta_1 &=& 
-  \frac{\tau_{\rm rel}}{1 - i \omega_c \tau_{\rm rel}}\frac{T}{{\cal N}_{\rm ph}} I_1 I_2 \ ,
\\ \label{xi2}
\zeta_2 &=&   
 \frac{\tau_{\rm rel}}{1 - i \omega_c \tau_{\rm rel}}\frac{T}{{\cal N}_{\rm ph}} I_2^2 \ ,
\\ \label{xi3}
\zeta_3 &=&  
\frac{\tau_{\rm rel}}{1 - i \omega_c \tau_{\rm rel}}\frac{T}{{\cal N}_{\rm ph}} I_1^2 \,,
\ea
and therefore $\zeta_1^2= \zeta_2 \zeta_3$, meaning that one of the relation for positive entropy production is saturated.

For $\omega_c=0$ and in the non-relativistic limit we recover the expressions of the bulk viscosity coefficients  that were obtained by Khalatnikov~\cite{IntroSupe}.
For non-vanishing frequencies we obtain bulk viscosities that have both real and imaginary parts,
as it occurs when there are periodic perturbations in time in the system, see \cite{landaufluids}.
For the computation of the dissipated energy in the system, only the real part matters (see
for example \cite{Lindblom:2001hd}). One can see  that the various coefficients attain their maximum value when $\omega_c = \frac 1{\tau_{\rm rel}}$. 

At this point it is easy to show  that for phonons with a linear dispersion law all the bulk viscosity
coefficients vanish, independent of whether the system is conformal invariant or not, see the Appendix~\ref{Linearlaw} for a short proof. This is in agreement with the
results obtained for $^4$He~\cite{Khalat-Cherni}, and for cold Fermi atoms close to the unitary limit~\cite{Escobedo:2009bh}.
Then we consider the first correction to the phonon dispersion law given by the cubic term, in the form given by
Eq.~(\ref{cubic-law}). After defining  the dimensionless parameter
\be 
\label{x-variable}
x =\frac{B T^2}{c_s^3} \,,
\ee
we compute the various quantities evaluated with the equilibrium phonon distribution function to the leading order in $x$. 
In this way we obtain that the number of phonons per unit volume is given by
\be\label{N-ph}
{\cal N}_{ph}  = \frac{T^3 }{2\pi^2 c_s^3}\Big(\Gamma(3) \zeta(3)- x\,  \Gamma(6)\zeta(5) + {\cal O}(x^2)\Big)\,,
\ee
while the  entropy  turns out to be
\be\label{entropy-exp}
 S_{ph} =   \frac{T^3}{6 \pi^2 c_s^3}\Big(\Gamma(5) \zeta(4) -3 x \Gamma(7)\zeta(6)+  {\cal O}(x^2)\Big) \,,
\ee
where $\Gamma(z)$ and $\zeta(z)$ stand for the Gamma and Riemann zeta functions, respectively.

For the evaluation of the functional derivatives that appear in Eq.~(\ref{defIs}) we use the same
strategy followed in Ref.~\cite{Escobedo:2009bh}. We use as independent variables the temperature $T$
and the zero temperature chemical potential~\cite{footnote}. Using a Maxwell relation, and the phonon contribution to the
chemical potential, we arrive at the expressions (see Ref.~\cite{Escobedo:2009bh} for more details)
\be\label{I1-ext}
I_1 = \frac{60}{7 c_s^7 \pi^2} 
 T^5 \Big(\pi^2 \zeta(3) - 7 \zeta(5)\Big) \left(c_s \frac{\partial B}{\partial \bar n} - B \frac{\partial c_s}{\partial \bar n}\right) \,,
\ee
and
\be\label{I2-ext}
I_2 = -\frac{20}{7 c_s^7 \pi^2} 
 T^5 \Big(\pi^2 \zeta(3) - 7 \zeta(5)\Big) \left(2 B c_s + 3\bar n \left(c_s \frac{\partial B}{\partial \bar n} - B \frac{\partial c_s}{\partial \bar n}\right)\right)\,.
\ee

For a conformal invariant non-relativistic superfluid one expects that $\zeta_1$ and $\zeta_2$ vanish, while there is no constraint
on $\zeta_3$~\cite{Son:2005tj}. For the ultracold Fermi superfluid in the unitarity limit this is what one indeed finds 
\cite{Escobedo:2009bh}.
One can expect that these results hold for relativistic superfluids, as well. In the following we shall explicitly prove that this is what happens.

Let us study under which conditions in a conformally invariant system the phonon contribution to the first and second   bulk viscosity
coefficients vanish. From the expressions in Eqs.~(\ref{xi2}) and (\ref{xi3})  
this is equivalent to asking that $I_2=0$. In a conformally invariant system the speed of sound is fixed,  $c_s^2= 1/3$. Further, since $\mu_0$ is the only energy scale it follows that by dimensional analysis $B \propto c_s/\mu_0^2$ and $\bar n \sim \mu_0^3$. Upon considering these scaling behaviors  in Eq.~(\ref{I2-ext}), and in the case that $B= a c_s/\mu_0^2$ one finds that $I_2$ vanishes.

In Section~\ref{bulkcfl} we will discuss the CFL superfluid, where we  find that for massless
quarks, and neglecting the running of the gauge coupling constant,
 the parameter $B$ has the dependence expected in a conformal invariant system, and thus  the only non-vanishing
bulk viscosity turns out to be $\zeta_3$.

\section{Zero-frequency temperature dependence of
relaxation time and bulk viscosities of a cold relativistic superfluid}
\label{Scalingbulk}

In the previous Section we have derived the expressions of the three independent bulk viscosity coefficients. In order to determine their actual temperature dependence we now estimate the temperature dependence of the relaxation time for the pertinent process. We shall consider the situation where $\omega_c =0$; the non-static case  can be easily deduced as well. For the time being the only relevant assumption we will make is that the contribution of
other quasiparticles to transport phenomena is thermally suppressed.  Therefore, we shall restrict to   the situation where $T$ is much less that any  energy gap of the system.
 
The computation of the phonon contribution to the bulk viscosities presents different subtleties.
As we saw in Section~\ref{EFF}, the phonon self-interactions are described by a low energy effective
field theory, whose form is the same for all the systems that share the same global symmetries.    
However, the coefficients that appear in the low energy Lagrangian depend on the system considered, as their
values have to be matched with the corresponding microscopic theory. As we saw, the bulk viscosities turn out to
depend on the coefficient $B$ that appears in the phonon dispersion law, see Eqs.~(\ref{I1-ext}) and (\ref{I2-ext}).
The transport coefficients also depend on the relaxation time of the collisional process responsible of
the corresponding dissipative phenomena. For bulk viscosity those are collisions that change the number
of phonons.  It turns out that the sign of $B$ decides whether some collisional
processes are kinematically allowed or not, and ultimately, this affects how the relaxation time associated to bulk viscosity scales with $T$.

For a dispersion law where $B$ is positive, the so-called Beliaev process \cite{Beliaev}
which describes the decay of one phonon into two,
$\varphi \rightarrow \varphi \varphi$, is
kinematically allowed. The decay rate associated to this process is given by
\be
\Gamma_{\rm ph} =  \int dP\, dK \, dQ \,
| {\cal M}(P,Q,K)|^2  n_{\rm eq}(p_0) \left(1 +  n_{\rm eq}(q_0)\right)\left(1 +  n_{\rm eq}(k_0)\right) (2 \pi)^4 \delta^{(4)} (P-K-Q) \ ,
\ee 
where $P^\mu = (p_0, {\bf p})$  is the four momentum and we have defined the integration measure as
\begin{equation}
\int dP \equiv \int \frac{d^3 p}{(2 \pi)^3} 2 \Theta(p_0) \delta(p_0^2 - \epsilon_p^2) \ ,
\end{equation}
and  $n_{\rm eq}$ is the Bose-Einstein equilibrium distribution function. The squared of the scattering amplitude at LO is expressed as
\be
| {\cal M}(P,Q,K)|^2 = g_3^2 \left(p_0 \, Q \cdot K + q_0 \, P \cdot K + k_0
\, Q \cdot P \right)^2 \,,
\ee
written in terms of the momenta of the phonons and of the three phonons self-coupling constant defined in Eq.~(\ref{LO-coupl}).
Even without an explicit evaluation of the above integral, one can infer that 
$ \Gamma_{\rm ph} \sim g_3^2 T^{8}/c_s^6$, and that the relaxation time scales with the temperature as
$\tau_{\rm rel} \sim c_s^3/g_3^2T^{5}$.  

As we have already shown in the previous Section, for a conformally invariant system, $\zeta_1=\zeta_2=0$, whereas $\zeta_3\neq 0$.
Upon substituting the scaling behavior of $\tau_{\rm rel}$, of $I_1$ and $I_2$ given in Eq.~(\ref{I1-ext}) and (\ref{I2-ext}), of ${\cal N}_{ph}$ given in Eq.~(\ref{N-ph}), in Eqs.~(\ref{xi3}) one finds that  $\zeta_3 \sim T^3$.  When conformal symmetry is broken, it follows that the three independent bulk viscosity coefficients do not vanish and have the same temperature dependence, {\it i.e.} $\zeta_i \sim T^3$.

Including NLO corrections both in the dispersion law $\epsilon_p$, and in  the self-coupling $g_3$, corrects
this leading order behavior with terms of order
$B^2 T^2/c_s^3$, which at low temperatures are negligible.

\begin{center}
\begin{figure}[h!]
\includegraphics[width=3.in,angle=0]{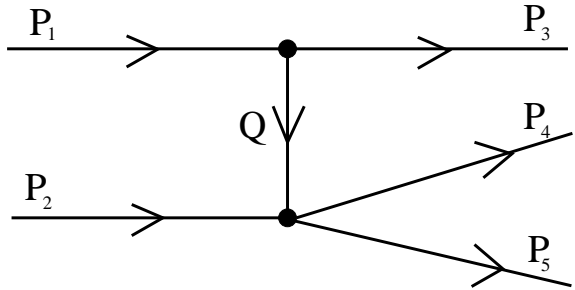}
\caption{ One of the Feynman digrams for the process $\varphi \varphi \rightarrow \varphi \varphi \varphi$. The scattering matrix  of this process is  reported in Eq.~(\ref{scattering}).
 } \label{Feynman}
\end{figure}
\end{center}

For a phonon dispersion law that curves downward, $B < 0$, the  Beliaev process is not kinematically allowed, and 
one has then to consider the five phonons collisional  processes. The decay rate for these processes is very different with respect to the $\varphi \to \varphi \varphi$ process. Further, as we discuss below, collisions with large or small scattering angle also have a rather different decay rate.

The decay rate of the process $\varphi \varphi \rightarrow \varphi \varphi \varphi$ is given by
\be
\Gamma_{\rm ph} =  \int \prod_{i=1, ..,5} dP_i
| {\cal M}_{5 \varphi}|^2  n_{\rm eq}(p_0^1)n_{\rm eq}(p_0^2)
 \left(1 +  n_{\rm eq}(p_0^3)\right)\left(1 +  n_{\rm eq}(p_0^4)\right) \left(1 +  n_{\rm eq}(p_0^5)\right)(2 \pi)^4 \delta^{(4)} (P_1+P_2-P_3-P_4-P_5) \, ,
\ee 
where ${\cal M}_{5 \varphi}$ is the scattering matrix for the process considered.

After a careful analysis, one realizes that this rate  has a very different behavior if one considers large or small angle
collisions. Let us consider the Feynman diagram reported in Fig.~\ref{Feynman}. The square of the
scattering matrix behaves as
\begin{equation}
| {\cal M}_{5 \varphi}(P_1,P_2,P_3,P_4,P_5)|^2 = g_3^2 g_4^2 \frac{F_3^2(P_1,-P_3,-Q) F_4^2(P_2, Q,-P_4,-P_5)}{\left(q_0^2 - \epsilon_q^2
\right)^2} \, , \label{scattering}
\end{equation}
where $Q^\mu= (q_0, {\bf q}) = (p_{10}-p_{30},{\bf p}_1 - {\bf p}_3)$ is the momentum transfer, and $F_3$ and
$F_4$
are polynomial of the phonon momenta that we shall  not specify here, but that can be easily
inferred from the Lagrangian~(\ref{L-BGB}). 
If one considers a collision with large angle scattering, naive power counting suggests that 
$ \Gamma_{\rm ph} \sim g_3^2 g_4^2 T^{16}/c_s^{12}$, and thus $\tau_{\rm rel} \sim c_s^9/(g_3^2 g_4^2T^{13})$, which turns out to be
a  rather large value for low temperatures. 

Processes associated to  small angle scattering behave very
differently. Let us start by considering phonons with a linear dispersion law. Then, the scattering
matrix associated to a small angle collision behaves as
\begin{equation}
\label{5ph-coll}
| {\cal M}_{5 \varphi}(P_1,P_2,P_3,P_4,P_5)|^2  = g_3^2 g_4^2 \frac{F_3^2(P_1,-P_3,-Q) F_4^2(P_2, Q,-P_4,-P_5)}
{4 c_s^4 p_1^2 p_3^2 (1 - \cos{\theta_{13}})^2} \ ,
\end{equation}
where $\theta_{13}$ is the angle between the vectors ${\bf p}_1$ and ${\bf p}_3$. In the limit where
 $\theta_{13} \rightarrow 0$ this process presents a 
collinear singularity, which needs to be regularized.
 This might be done by considering 
a phonon dispersion law that includes a cubic term, as then in the denominator of Eq.(\ref{5ph-coll}) one
has a term proportional to $B^2$, which does not vanish in the $\theta_{13} \rightarrow 0$ limit.

Let us stress here that while we have based our discussion on the Feynman diagram in Fig.~\ref{Feynman}, there
are other channels for five phonons collisions that also present similar collinear singularities.
The naive power counting associated to small angle collisions then suggests that
$\Gamma_{\rm ph} \sim g_3^2 g_4^2 T^{12}/(c_s^6 B^2)$,
and thus a relaxation time scaling as $\tau_{\rm rel} \sim c_s^3 B^2/(g_3^2 g_4^2T^9)$, which is a much faster process than that corresponding
to  large angle collisions. Because in the computation of the transport coefficients one has to consider first the
collisions with the shorter relaxation time, one then concludes that in this case, and when conformal symmetry is broken,
all the bulk viscosity coefficients scale with the temperature as $\zeta_i \sim \frac 1T$. Corrections to this leading behavior would be very suppressed, as discussed earlier.

Summarizing, bulk viscosities in the cold regime of relativistic superfluid strongly depend on the form of the phonon dispersion law. If conformal symmetry is unbroken, one should have $\zeta_1 = \zeta_2 =0$,
while the three bulk viscosity coefficients are non-zero when conformal symmetry is broken.
When $B>0$ the process $\varphi \to \varphi\varphi$ is kinematically allowed and gives the dominant contribution to bulk viscosity. In this case one may expect a rather small value for the bulk viscosities, with scaling  $\zeta_i \sim T^3$.
When $B<0$ one has to consider the processes
$\varphi \varphi \to \varphi\varphi\varphi $ which is  
 dominated by small angle scatterings. These processes give  
 scalings  $\zeta_i \sim 1/T$. 
Let us finally remark that
while our discussion has been done for relativistic superfluids, it  is also valid for non-relativistic superfluids.

\section{Bulk viscosities for the CFL phase due to superfluid phonons}
\label{bulkcfl}

At asymptotically high densities  quark matter is  believed to be in the color-flavor locked  phase~\cite{Alford:1998mk}. In this phase up, down and strange  quarks of all three colors form  zero-momentum spinless Cooper pairs. 
The CFL pairing pattern spontaneously breaks both the flavor chiral and the baryonic number $U(1)_B$
symmetries, leaving a discrete $Z_2$ symmetry. There are eight (pseudo) Goldstone modes associated to the chiral group, and one
exact massless Goldstone boson associated to the breaking of the $U(1)_B$, which is responsible for the superfluid properties of the system. All gluons are massive, and all quarks are gapped. All the properties of CFL quark matter can be computed from
QCD in the high density limit~\cite{reviews}. 

In this Section we derive the bulk viscosity coefficients of CFL quark matter using the strategy discussed in the first part of the manuscript.

\subsection{Conformal limit}

As shown in Section~\ref{EFF}, the effective low energy Lagrangian for the Goldstone boson
can be obtained from the pressure \cite{Son:2002zn}, and for the   CFL quark matter at asymptotic densities we have that 
\begin{equation}
 \label{freeEOS} 
P [\mu_0]
=\frac{3}{4\pi^2} \mu_0^4 \ ,
\end{equation}
where $\mu_0$ is the quark chemical potential. At very high $\mu_0$,
where the coupling constant is small $g (\mu_0) \ll 1$, the effects
of interactions and the effects of Cooper pairing are subleading
and have been neglected in Eq.~(\ref{freeEOS}). Also, at very high densities
one might neglect the quark masses,  as $m_q \ll \mu_0$.

The values of all the coefficients which appear in the phonon Lagrangian at LO, Eq.~(\ref{L-BGB}), can now be extracted as explained in Section~\ref{EFF}, and   are given by
\begin{equation}
\label{parameters}
c_s^2  =  \frac 13  \ ,  \qquad
g_3  =  \frac {\pi}{9 \mu_0^2} \ , \qquad
g_4  =  \frac {\pi^4}{108 \mu_0^4} \ .
\end{equation}

It is possible to compute the phonon dispersion law at NLO for massless quarks by matching with QCD
at finite density ~\cite{Zarembo:2000pj,foot}. In this way one obtains that
\be\label{disperion-law}
\epsilon_p = c_s p   \left( 1 - \frac{11}{ 540} \frac{p^2}{\Delta^2} + {\cal O}\left(\frac{p^4}{\Delta^4}\right)\right) \,,
\ee
where $\Delta$ is the CFL gap. In the asymptotic high density limit, the gap can be computed from QCD 
~\cite{Son:1998uk}
\be
\Delta \simeq \mu_0 g^{-5} \exp\left(-\frac{3 \pi^2}{\sqrt{2} g}\right)b_0 \ ,
\ee
where $g$ is the QCD gauge coupling constant, and $b_0 = 512 \pi^4 (\frac 23)^{5/2} \exp{ \left(-\frac{\pi^2+4}{8} \right)}$ 
\cite{Brown:1999aq,Wang:2001aq}.

For sufficiently large chemical potentials, one can neglect the running of the coupling constant
 \cite{Schafer:2004yx}.
  Then one can consider that the gap  $\Delta \propto \mu_0$, that is,
the gap only depends linearly on the chemical potential and therefore does not break conformal invariance.
 Moreover in the regime we consider the total particle density is $\bar n \approx \frac{3 \mu_0^3}{\pi^2}$,
consistent with the approximation to the CFL pressure, and being in the regime $T \ll \mu_0$.
Since $\bar n \propto \mu_0^3$ and $\Delta \propto \mu_0$, it follows that 
\be 
I_1 =   -\frac{60}{7 c_s^7 \pi^2} 
 T^5 \Big(\pi^2 \zeta(3) - 7 \zeta(5)\Big)  \frac{2 B c_s}{3 \bar n} \ , \qquad
  I_2 = 0 \, , \label{I1-I2}
\ee 
where, according with Eq.~(\ref{disperion-law}), the coefficient of the cubic term in momentum is given by 
 \be 
B= -\frac{11 c_s}{540 \Delta^2} \,.\ee

Since  $B < 0$, we have to consider that the processes giving the  leading contributions
to the bulk viscosity are small angle five phonons collisions.
The estimate of the relaxation time for these processes is given in Sec.~\ref{Scalingbulk} and turns out to be
\begin{equation}
\tau_{\rm rel} \sim 
\frac{c_s^3}{T^9}\frac{\mu_0^{12}}{\Delta^4} \ .
\end{equation}

We consider first  the case $\omega_c=0$. From Eqs.(\ref{I1-I2})
 we conclude that $\zeta_1=\zeta_2=0$,  while   the third bulk viscosity does not vanish and depends parametrically on the physical scales of the problem as
\begin{equation}
\zeta_3 \sim \frac{1}{T} \frac{\mu_0^6}{c_s \Delta^8} \, .
\end{equation}
We remark that 
 these are only  approximated results that arises in the $g \ll 1$ limit, after neglecting the running of the QCD gauge coupling constant and the effect of the strange quark mass.
These are good approximations  in the regime of high density~\cite{Schafer:2004yx}. 
For non-vanishing values of $\omega_c$ one can accordingly find the value of the only non-vanishing transport coefficient, given the expressions
of the relaxation time and zero-frequency viscosity, see Eq.~(\ref{xi3}).

\subsection{Including scale breaking effects due to the strange quark mass}

The bulk viscosities are sensitive to scale breaking effects.
The quantum scale anomaly breaks the conformal symmetry introducing, through dimensional transmutation, the quantum scale $\Lambda_{\rm QCD}$.
One could then compute 
$g$-corrections to the pressure of quark matter, which would then modify the different
coefficients of the phonon effective field theory, introducing terms proportional to the QCD beta function.  
In the very high $\mu_0$ limit, when $g (\mu_0) \ll 1$, we expect this to be a rather negligible
effect, as mentioned in the previous Section. Also consider that for large values of the chemical potential the gap parameter, $\Delta$, does not break scale invariance because  in this case  one has   $\Delta \propto \mu_0$.

The inclusion of quark mass effects leads to the breaking of scale invariance. 
In the CFL phase, the three light quarks participate in the pairing process,
and therefore the largest effect comes from the  strange quark mass $m_s$. 
However, one cannot consider arbitrarily large values of the strange quark mass, because otherwise the CFL phase becomes chromo-magnetically unstable~\cite{Casalbuoni:2004tb}. Therefore, we shall take 
$m_s^2 < 2 \Delta \mu_0$,  which is the threshold value under which the CFL phase is stable. Since $m_s \ll \mu_0$ we will consider  the leading order corrections  in $m_s^2/\mu_0^2$. After imposing the constraints of electrical neutrality and beta equilibrium of quark matter,
the first correction of the order $m_s^2/\mu_0^2$ to the pressure reads~\cite{reviews}
\begin{equation}
P [\mu_0]
=\frac{3}{4\pi^2} \left(\mu_0^4 - \mu_0^2 m^2_s \right) \, .
\end{equation}

Using Eqs.~(\ref{cs}) and (\ref{LO-coupl}) we find that to the order $m_s^2/\mu_0^2$ the coefficients of the LO phonon Lagrangian
are corrected due to this scale breaking effect as
\be
\label{m-speedsound}
c_s^2  =  c_{s,0}^2 \left( 1 - \frac{m^2_s}{3 \mu_0^2} \right) \ ,  \qquad
g_3  =  g_{3,0} \left( 1 + \frac{m^2_s}{4 \mu_0^2} \right) \ , \qquad
g_4  =  g_{4,0} \left( 1 + \frac{m^2_s}{3 \mu_0^2} \right) \, ,
\ee
where the $c_{s,0}$, $g_{3,0}$ and $g_{4,0}$   refer to the values obtained
when $m_s =0$.
Thus, the leading order effect of the strange quark mass in the phonon
Lagrangian is to modify the velocity of the phonon, that is the speed of
sound,  and a finite renormalization of the cubic and quartic phonon self-couplings.

The phonon dispersion law at cubic order including quark mass
corrections has not yet been computed in the literature, as it is subtle.
Because a mass term in the QCD Lagrangian couples the particle
and antiparticle degrees of freedom, that computation requires the knowledge of
the antiquark propagator in the CFL phase, exactly as it occurs in the computation of the CFL meson masses
 \cite{Son:1999cm,Son:2000tu,Manuel:2000wm}.  In any case, the correction of the coefficient of the cubic term in momentum in
the phonon dispersion law will be 
of the form
\begin{equation}\label{newB}
B = B_0 \left( 1 + {\bar b}_s \frac{ m^2_s}{\mu_0^2} \right) \,  ,
\end{equation}
with $\bar b_s$ a dimensionless constant. Since we are only interested in determining the scaling laws and  not the   numerical factors, we leave the computation of this constant for a future publication.  

Upon substituting the expression of Eq.~(\ref{m-speedsound}) and (\ref{newB}) in Eqs.~(\ref{I1-ext}) and (\ref{I1-ext}), one finds
\begin{eqnarray}
I_1 & = &   -\frac{60}{7 c_{s,0}^7 \pi^2} 
 T^5 \Big(\pi^2 \zeta(3) - 7 \zeta(5)\Big)  \frac{2 B_0 c_{s,0}}{3 \bar n_0} \left(1+
(2 {\bar b}_s +1)  \frac{m_s^2}{\mu_0^2} \right)
 \ , \\
  I_2 & = & -\frac{20}{7 c_{s,0}^7 \pi^2} 
 T^5 \Big(\pi^2 \zeta(3) - 7 \zeta(5)\Big)  2 B_0 c_{s,0} \left(
\frac 53 - {\bar b}_s  \right) \frac{m_s^2}{\mu_0^2} \,.
\end{eqnarray}

The scale breaking terms affect the relaxation times as well, because the phonon self-couplings are modified for $m_s \neq 0$, as shown in 
 Eqs.~(\ref{m-speedsound}). A naive power counting analysis 
suggests that
\be
\tau_{\rm rel} \sim \tau_{\rm rel,0} \left( 1 + (2 {\bar b}_s - \frac 13) \frac{m^2_s}{\mu_0^2} \right) \,.
\ee

We discuss now the values of the bulk viscosities for the $\omega_c =0$ case, the non-static case
is then easily deduced as well.
As we are only interested in the scaling laws associated to the bulk viscosity coefficients,
we conclude that to order $m_s^2/\mu_0^2$ 
\be
\zeta_1 \sim  \frac{1}{T} \frac{\mu_0^9}{c_{s,0} \Delta^8}  \frac{m^2_s}{\mu_0^2} \ ,
 \qquad  \zeta_2 \sim \frac{1}{T} \frac{\mu_0^{12}}{c_{s,0}\Delta^8} \frac{m^4_s}{\mu_0^4} \ .
\ee

Further, taking into account that $\Delta \propto \mu_0$,  one obtains the dependence of the different 
coefficients with the basic physical scales of the problem, $T, \mu_0, m_s$. In this way one finds
$[\zeta_1] = \frac{m^2_s}{T \mu_0}$,  $ [\zeta_2] = \frac{m_s^4}{T}$ and  $[\zeta_3] = \frac{1}{T \mu_0^2}$.
Note that the scaling behavior of $\zeta_2$ was previously found in
Ref.~\cite{Manuel:2007pz}. There the phonon Boltzmann equation was solved numerically, using a variational
method. In that article it was assumed a phonon spectrum linear in momentum, and the interactions were extracted from
the LO Lagrangian only, neglecting all phonon physics at NLO.  In the linearization of the transport equation, the variations of the speed of
the phonon with the total density were neglected, and that is why a non-zero result was obtained, even if the whole computation
was carried out with the LO phonon physics.

\section{Conclusions}
\label{conclu}

Superfluidity arises at low temperatures after the spontaneous breaking of a global $U(1)$ symmetry. It is possible to determine the  expression of the low energy   Lagrangian  of the Goldstone mode by using symmetry arguments. However, the coefficients of the effective field theory are not fixed  by the symmetries of the system; they   have to be matched with the microscopic field theory, and thus  depend on the physics at the very short scales. 

In this paper we have used the 
 effective field theory describing the low energy properties of a superfluid system, together with kinetic theory in the relaxation time approximation, to obtain 
the contribution of the superfluid phonon to the bulk viscosity coefficients. We have first presented a general discussion on the low energy Lagrangian of the system. Then, we have determined the scaling behavior of the bulk viscosities with $T$. In order to do this we have estimated the relaxation time of the processes responsible for bulk viscosity. The allowed processes  depend on the sign of
the   correction to the linear dispersion law of the phonon.   If the correction to the linear dispersion law curves upward, one phonon can decay into two, but this process is kinematically forbidden in the opposite case. If the correction to the linear dispersion laws curves downward, then the most important process is $\varphi\varphi \to \varphi\varphi\varphi $ which is dominated by small angle scatterings. We have presented a general discussion on how the relaxation times behave in both cases in Section~\ref{Scalingbulk}.

In the derivation of the transport coefficients we have neglected the contribution of other quasiparticles. This approximation is  justified in the situation where
$T$ is much less that any other energy gap of the system. If not, one should take into account the contribution of other quasiparticles as well.
 
We have used the general expressions of the transport coefficients for a relativistic superfluid to determine the phonon contribution to the bulk viscosity coefficients of  CFL quark matter.
 In order to do this we have determined the scaling of the relevant relaxation times. Let us stress that with the relaxation time approximation
we can only obtain the correct parametric behavior of the transport coefficients, but this should be enough to get their
correct scales to be used, for example, in the study of the r-mode oscillations of a quark star. Previous analysis for the r-modes
of a CFL quark star have been carried out in Refs.~\cite{Madsen:1999ci,Sa'd:2008gf,Jaikumar:2008kh}, but there the transport coefficients that
we have studied as well as the mutual friction that operates in rotating superfluids~\cite{Mannarelli:2008je} were ignored.  Let us also point
out that the knowledge of the bulk viscosities is relevant for the evaluation of the cavity conditions in quark stars~\cite{Madsen:2009tb}.

It would be  interesting to study  the phonon contribution to the
bulk viscosities of a neutron superfluid applying the general expressions derived in the present paper,
as that contribution has been ignored in astrophysical applications. 
It might also be interesting to study the phonon contribution to transport phenomena with holographic
techniques, as this should be independent of whether the underlying microscopic theory is strongly or weakly coupled.

\begin{acknowledgments}
We thank Felipe Llanes-Estrada and Krishna Rajagopal for useful comments.
This work has been supported by the Spanish grant
FPA2007-60275. MM thanks the Department of Physics of the University of Bari for the kind hospitality during the completion of the present work. 
\end{acknowledgments}

\appendix
\section{Phonons with a linear dispersion law}
\label{Linearlaw}
For phonons with a linear dispersion law one can write
\be
{\cal N}_{ph}  =  3 \frac{\Gamma(3) \zeta(3)}{\Gamma(5) \zeta(4)} S_{ph}  \,,
\ee
meaning that  ${\cal N}_{\rm ph} \equiv {\cal N}_{\rm ph}(S) $ and the phonon density is independent of $\bar n$
at this order. From Eqs.~(\ref{xi1},\ref{xi2},\ref{xi3}) it follows that  for phonons with a linear dispersion law $\zeta_1=\zeta_2=\zeta_3=0$. Note that this result is true for any form of the speed of sound,
$c_s \equiv c_s(\mu_0)$.  The fact that  for phonons with a linear dispersion law   ${\cal N}_{\rm ph}$ is independent of $\bar n$ can also be explicitly proven. By the chain rule we have that
\be
\frac{\partial {\cal N}_{ph}}{\partial \bar n} = \frac{\partial {\cal N}_{ph}}{\partial T}\frac{\partial T}{\partial \bar  n} + \frac{\partial {\cal N}_{ph}}{\partial \mu_0}\frac{\partial \mu_0}{\partial \bar n} \label{dn-dn}\,,
\ee
that we can rewrite as
\be
\frac{1}{3  {\cal N}_{ph}}\frac{\partial {\cal N}_{ph}}{\partial \bar n} =  \frac{1}T  \left( \frac{\partial T}{\partial \bar n}\right)_S -     \frac{1}{c_s}\left( \frac{\partial c_s}{\partial \bar n}\right)_S \,.
\ee
From the value of the entropy with a linear phonon dispersion law we have that 
\be
c_s = \left(\frac{\Gamma(5) \zeta(4)}{6 \pi^2}\right)^{1/3} \frac{T}{S^{1/3}}\,,
\ee
therefore 
\be \frac{1}{c_s}\left( \frac{\partial c_s}{\partial \bar n}\right)_S  =  \frac{1}T  \left( \frac{\partial T}{\partial \bar n}\right)_S \,,\ee
and it follows that $\frac{\partial {\cal N}_{ph}}{\partial \bar n} = 0$.


\begin{thebibliography}{99}
\bibitem{footnote}
As explained in Ref.~\cite{IntroSupe}, Eq.(18-13), the chemical potential receives a contribution from the thermal excitations of the system. We shall refer to the chemical potential that does not include such excitations as the zero temperature  chemical potential.
\end{thebibliography}
\end{document}